\documentstyle[12pt]{article}

\begin{document}

\pagestyle{empty}

\hsize 14.5 cm

\centerline{\Large{Can Economics Afford Not To Become  Natural
Science?}}

\vskip 1 cm

\centerline {Bikas K. Chakrabarti}
\centerline{Saha Institute of Nuclear Physics, Kolkata 700064, India}
\centerline {\&}
\centerline{Economic Research Unit, Indian Statistical Institute, 
Kolkata 700108, India} 

\medskip

\centerline {\small {E-mail: bikask.chakrabarti@saha.ac.in}}

\vskip 1 cm

\noindent
The moot question is: ``Can economics be a Physical Science?" I 
would rather address a more general question: ``Can economics 
or sociology avoid joining  Natural Science?"

The  answer of course depends on how we look at 
sociology or economics. Are they
going to remain prescriptive as part of ethics 
(as often in political economics, or
as social norms evolved through the history of 
the human kind) or they aspire to be
a meaningful ``tool" or ``model" to comprehend the 
social dynamics and be useful to
``engineer" the social dynamics when needed.

Epistemologically, our knowledge about truth can be 
either deductive or inductive.
Mathematics is an usual example of the deductive 
knowledge (though not all of it can
be deduced from axiomatic logic). Mathematical truths 
do not necessarily require any
laboratory test or supports from ``observations" of 
the ``nature" to prove or
validate them. Linguistically it is like the tautology 
``A bachelor does not have 
wife"; one need not check with each and every bachelor 
to confirm the statement --
first part of the sentence confirms the second. 
The same is true about the statement
``two plus two  equals  four". Mathematical truths 
are analytical truths, and 
mathematics therefore is not generally considered to 
be an empirically founded
science, though it has been historically considered 
as the ``mother' of all natural
sciences.  Mathematics is  employed as a condensed 
form of the (deductive) logic,
and as such, does not by itself reveal any new truth 
about nature (Whitehead and
Russell, 1910, 1912, 1913; see also Kneebone, 2001). Natural 
sciences, however,  are
basically inductive in origin, based on natural or 
laboratory observations.  
The statement ``The sun rises every twenty four 
hours on the east" is not a
tautology or analytical truth. Though east may be 
defined as the direction of
sun-rise, that it rises every twenty four hours, is 
an inductive (or empirically 
observed) truth, and therefore tentative (not like 
mathematical truths).  The tools
of mathematics and logic are employed to find and 
establish relationships among
these observations, sometimes with the help of a 
common (abstract) hypothesis about
the functioning of the nature (see e.g., Russell, 1967). 

Mathematics has played, and still plays a significant 
role in developing the
consistency logic in any natural science. This is 
particularly true in physics. In
chemistry and biology, its role is gradually being 
replaced by (linguistic) logic.
Even in physics, there are increasing replacements 
by (gate based) computer
algorithms or Boolean logic. There are even
arguments that a new kind of (natural) science will 
be developed soon and will
replace all our mathematics and linguistic logic 
by computer
algorithms (Wolfram, 2002).  Even when the mathematics 
is employed as the major tool
in developing any natural science, as in  physics, it 
is considered as applied
mathematics; it does not generally contribute to the 
development of mathematics, as
done by the mathematicians. Hence, even mathematical 
physics need not be confused
with mathematics itself! Novel applications of 
mathematics do not lead to any
success in physics unless it builds on empirical facts 
and leads to predict or
engineer other empirical facts. Also, approximate 
mathematical solutions in the
intermediate steps to comprehend the nature are most 
welcome (see e.g., Stanley,
2013; Weatherall, 2013). Even at its best, such 
intermediate analysis is just applied
mathematics, and not (pure) mathematics. The same 
should be true for mainstream
economic theories today. They do not by themselves 
contribute to the development of
mathematics or to the development of economics as
 science, unless these
(applied) mathematics comes at the intermediate 
stage to take us to the predictable set
of empirical truths from the (base) set of 
observed truths. 

Major use or
application of one field to another 
never leads to identity: In any concert, 
many instruments are used,
including the sound system, which is common. This 
never suggests identification of
the concert with the sound system! More seriously 
perhaps,  most concerts employ a
piano or a  harmonium whose players play key roles. 
Yet, the main conductors of the
concerts may not be those players, nor even players of 
those instruments! Applied
mathematicians need not be identified either as 
(natural) scientists or as (pure)
mathematicians! These mistakes can be costly: 
``I suspect that the attempt to
construct economics as an axiomatically based hard science 
is doomed to fail" (Solow,
1985).

These quantifiable links between the seemingly often unrelated
observations help to identify the generic truths of nature 
and form the basis for
comprehension of the scientific (observational) laws. 
However, the perception and
comprehension of truth can be partial and incomplete. 
A proverbial Indian story
(Buddhist Udana) depicts a few blind people touching 
different parts of an elephant:
the trunk, tusk, leg, tail etc., and interpreting 
them as different animate or
inanimate objects depending on their own perceptions, 
ideas or experience.
Physicists, chemists, biologists, all tend to do the 
same. Economists or
sociologists tend to make the same mistake more 
dramatically! In all its various
manifestations, inanimate, biological or sociological, 
mother nature perhaps employs
the same elegant truth code, the gene, which gets
suppressed partially and expressed differently in the 
various parts of her body.
Scientists having different perspectives in mind 
perceive them differently. Mother nature hardly cares 
whether we call them  physics or chemistry or biology, or
for that matter, economics or sociology. The
generic truth established therefore in one branch of 
natural science (say, in
physics) should not be invalid in another (say, in 
chemistry or biology) and the
same should hold true for economics if viewed as 
another branch of natural science
(Chakrabarti, 2013).

As argued earlier, mathematics in itself can not lead 
to any satisfactory theory or 
model of any natural system, unless the theory or the 
models are based on hypotheses
connected to some observations. Even approximate and 
tentative attempts to connect
various observations, employing
(deductive) logic or mathematics can lead eventually 
to a successful theory or
understanding about the nature. And that can be utilised 
to predict or suggest
desired outcomes in different contexts, leading to 
successful and much desired
engineering. Natural sciences therefore start with 
observations and end in
observations; in the middle it grows in successive 
stages as the tentative models
and theories try to accommodate more and more of the 
observations or data in more
elegant and comprehensive way. Indeed, ``lots 
of fields use mathematical models to
understand the world. But physicists have a 
particular way of thinking about
approximation and idealization. To make progress 
on interesting problems, physicists
always have to make assumptions and 
approximations" (Weatherall, 2013), allowing
respective engineering. In my mind, the mainstream 
economics is paying a heavy
price of making this mistake, by refusing to 
address the social engineering
problems! There are of course some skepticisms regarding
the success of mechanical models of individual's economic
activities, arguing that the economic agents, unlike particles,
incorporate the anticipated future prospects of their move into 
their dynamics. This argument, however, does not
stand on closer scrutiny: These anticipated prospects can
in fact be formally incorporated in their utility measures, 
which each one intends to maximize for future returns. Many 
of the specific forms 
of the utility  functions on the other hand
have been identified as precise equivalents of the 
thermodynamic entropy related to the collective dynamics (see
e.g., Chakrabarti et al., 2013), and the entropy maximization 
gives the time arrow or direction of the future in the collective 
dynamics. Another important point 
against such social engineering, some of which 
had been attempted earlier, comes from the adaptive change in the
behaviour of agents in response to the past attempts (of 
such engineering). This is
a more difficult problem, though not insurmountable: Hopfield-like
adaptive brain models are now quite established in science and technology
(see e.g., Hertz et al., 1991) and similar adaptive social learning
models for Minority Games and similar collective adaptive problems 
are being formulated 
and studied extensively these days (see e.g., Chakraborti et al., 2015).

Economics, apart from its inappropriate wisdom to become a hard science 
with exclusive axiomatic or mathematical foundations, also suffers from its 
seemingly immediate practical ambition of policy prescriptions
under the garb of political economy. In absence of any real science
of economics which can be implemented for any kind of social 
engineering, this high-handed prescriptions create even more 
problems.  The historical origin of the concept of inviting the 
involvement of politics or the government in individual's 
economic activites derives from the mercantile experience in the
colonial days. British merchants, who would even set prices
on their own terms in colonies like India, observed growing
instabilities or unemployement in their home market in years 
whenever there was a net trade deficit (export less than 
import, leading to a net outflow of gold). Mercantiles prescribed
involvement of the government (third party) to introduce taxes on 
import to control the loss of effective demand resulting from the
``laissez-faire"  of otherwise free individual traders or
economic agents. Political
economy was born to adovocate such  appropriate measures. Later,
this was further strengthened by the ``Keynesian  prescription" 
(Keynes, 1936) of fiscal measures by the government to curb the
loss of effective demand or slump  and the increased unemployment.  
In absence of any real foundation, one does not know when such
prescriptions work and when not (see e.g., Harcourt and Kriesler, 2013). 
Even with such great insights and intentions, economics has failed 
so far to comprehend the social instabilities. The appointments 
of powerful finance
ministers (with completely different level of assignments for 
the science ministers, who can never participate at any 
such root level)
  by the respective governments perhaps further worsen
the situations all over the world! 
These also suggest some major reviews are needed  for developing
``science of economics" urgently.

As mentioned above, mainstream economics today 
hardly cares for the subject's 
success in applications or social engineering. This has led to
the recent upsurge of some heterodox approaches, including
Econophysics.
Econophysics (term coined in 1995 in a conference 
on statistical physics in Kolkata; see e.g., Rosser,
2008;  Stanley, 2013)
views the problems of economics in a physical way 
and attempts to solve them
employing the techniques of physics. Indeed, 
viewed more generally, it tends to
believe that the dynamical aspects of the societies 
and of the markets in particular
are purely physical in origin and nature, in contrast 
to the academic belief in
economics that the complete economic world can be 
understood based on axiomatic
logic. 

Mark Twain (1883) noted long back ``There is
something facinating about science. One gets such
wholesale returns of conjecture out of such a trifling 
investment of fact." 
The knowledge and truths of social sciences like 
economics can only be fact-based or
inductive like in physics or other natural 
sciences, where deductions or logical
derivations, correlating various empirical
observations, are essential for their comprehension, 
organisation and growth. But that
need not be mistaken to be an indication for an 
essentially deductive science, where
logical elegance or mathematical beauty can 
dictate the truth! Just like biophysics
or biochemistry (each borrowing established knowledge 
i.e. truths from the parent
subjects like physics or chemistry to develop biology), 
econophysics can help to develop
economics utilizing the knowledge borrowed from 
physics which, being the oldest and
most established of all the natural sciences, 
can offer a major helping hand. Hence
the inevitability of econophysics.

Auguste Comte (1856) had already advocated for ``Social Physics"
more than a century and half ago. Pioneering arguments in favour 
of modelling social  sciences in the mould of physical sciences 
have recently been reiterated forcefully by several scientists: 
See for example, Schelling (1971), Galam et al. (1982), 
Mantegna and Stanley (1995, 1999), 
Stauffer et al. (2006),
Kirman (2010) and Galam (2012).  
Indeed there are several recent comments and discussion-notes
written by economists and social scientists, clearly saying 
``No" in response  to the question posed in the 
title of this paper: ``Ongoing work 
inspired by statistical physics shows that relatively simple 
models with plausible behavioural rules have the potential to 
replicate key empirical regularities of financial markets", wrote
Lux and Westerhoff (2009) on commenting about the failures of
mainstream economics in comprehending the 2007-2008 global
economic crisis, while in responding to some criticisms
on the statistical methodolgies of economics experiments, the use of 
which are growing 
extremely fast in the current economic theory literature, a
recent Editorial Note (2016) entitled `A far from dismal outcome'
 in The Economist writes ``Natural 
scientists may have to stop sneering at their economist 
brethren, and recognise that the dismal science is, indeed, a 
science after all."

\bigskip
\noindent Acknowledgement: I am grateful to Arnab Chatterjee,
Serge Galam, Alan Kirman,  
Thomas Lux, Satya  Majumdar, Richard M. Neumann, Purusattam Ray,
Parongama Sen and Dietrich Stauffer  for 
several important comments and critcisms on  this paper. 
\vskip 1.0 cm

\leftline {References:}

\medskip
\noindent B. K. Chakrabarti (2013), in Encyclopedia of Philosophy
and the Social Sciences, Ed.
B. Kaldis, Vol. 1, pp. 229-230, Sage Publications, Los Angeles

\medskip 
\noindent  B. K. Chakrabarti, A. Chakraborti, S. R. Chakravarty 
and  A. Chatterjee (2013), Econophysics of Income and Wealth Distributions, 
Cambridge University Press, Cambridge 

\medskip

\noindent A. Chakraborti, D. Challet, A. Chatterjee, M. Marsili, 
Y.-C. Zhang and B. K. Chakrabarti (2015), Physics Reports, Vol. 552, pp. 1-25

\medskip
\noindent A. Comte (1856), Social Physics: From the Positive 
Philosophy of Auguste Comte, Calvin Blanchard, New York

\medskip
\noindent Editorial Note (2016), The Economist, March 5th, 2016

\medskip
\noindent S. Galam (2012), Sociophysics: A Physicist's Modeling of 
Psycho-polotical Phenomena, Springer, Heidelberg

\medskip
\noindent S. Galam, Y. Gefen and Y. Shapir (1982), Mathematical 
Journal of Sociology, Vol. 9,  pp. 1-13

\medskip
\noindent G. C. Harcourt, P. Kriesler (Eds) (2013),
The Oxford Handbook of Post-Keynesian Economics, Vol. 2 (Critiques
and  Methodology), Oxford University Press, Oxford

\medskip
\noindent  J. A. Hertz, A. S. Krogh and R. G. Palmer (1991), Introduction 
to the Theory of Neural Computation, Santa Fe Institute Series, Westview 
Press, Colorado

\medskip
\noindent J. M. Keynes (1936), General Theory of Employment, Interest
and Money, Palgrave Macmillan, London

\medskip
\noindent A. Kirman (2010), Complex Economics, Routledge, New York

\medskip
\noindent G. T. Kneebone (2001), Mathematical Logic and the Foundations of
Mathematics: An Introductory Survey, Dover Publications, New York

\medskip

\noindent T. Lux and F. Westerhoff (1999), Nature Physics, Vol. 5, pp. 2-3

\medskip
\noindent R. N. Mantegna and H. E. Stanley (1995), Nature, Vol. 376, pp.
46-49 

\medskip
\noindent R. N. Mantegna and H. E. Stanley (1999), An Introduction to 
Econophysics: Correlationa and Complexity in Finance, Cambridge University
Press, Cambridge

\medskip
\noindent J. Rosser, Jr. (2008), in The New Palgrave Dictionary of 
Economics (Eds. S. N. Durlauf and L. E. Blume), Vol. 2, pp. 729-732, Palgrave 
Macmillan, New York

\medskip
\noindent B. Russell (1967), The Problems of Philosophy, Oxford University
Press, Oxford

\medskip
\noindent R. M. Solow (1985), American Economic Review (American Economic
Association), vol. 75, pp. 328-331

\medskip

\noindent T. C. Schelling (1971), Journal of Mathematical Sociology, Vol. 1,
pp. 143-186

\medskip
\noindent H. E. Stanley (2013), Interview by K. Gangopadhyay,
 IIM Kozhikode Society and
Management Review (Sage Publications), Vol. 2, pp. 73-78

\medskip
\noindent  D. Stauffer, S. M. de Oliveira, P. M. C. de Oliveira, 
J. S. de Simoes Martins (2006), Biology, Sociology, Geology by 
Computational Physicists, Elsevier, Amsterdam

\medskip

\noindent M. Twain (1883), Life on the Mississippi, Chatto and Windus, 
London 

\medskip
\noindent J. O. Weatheral (2013),  APS News (American Physical Society), 
Vol. 22(3), The Back Page

\medskip

\noindent A. N. Whitehead and B. Russell (1910, 1912, 1913), 
Principia Mathematica, Vols. I, II, III,  Cambridge University
Press, Cambridge

\medskip 

\noindent S. Wolfram (2002), A New Kind of Science, Wolfram Media, Illinois

\end{document}